# Topological Analysis and Synthesis of Structures related to Certain Classes of *K*-Geodetic Computer Networks

Carlos E. Frasser

A fundamental characteristic of computer networks is their topological structure. The question of the description of the structural characteristics of computer networks represents a problem that is not completely solved. Search methods for structures of computer networks, for which the values of the selected parameters of their operation quality are extreme, have not been completely developed. The construction of computer networks with optimum indices of their operation quality is reduced to the solution of discrete optimization problems over graphs. This paper describes in detail the advantages of the practical use of *k*-geodetic graphs [2, 3] in the topological design of computer networks as an alternative for the solution of the fundamental problems mentioned above which, we believe, are still open. Also, the topological analysis and synthesis of some classes of these networks have been performed.

## 1. The Problem of Generating Computer Networks with Optimum Topological Structure.

The urgent need for the construction of automated control systems, the great increase of the number of personal computers, and the rapid development of computer technology, which coincided with the colossal development of communications based on the use of digital channels, fiber optics, and cosmic technology, opened the path to the computer network development. The interaction of computers located over long distances, as a rule, is done through data transmission processors located in the main centers of information processing and packet interchange. The network of such centers is topologically represented by a strongly connected graph of diameter four and with an enormous number of vertices which, due to the demands of obtaining high indices (characteristics) of the network operation quality, must have a special structure [2, 5]. In the topological design early stage of computer networks is performed their *systemic analysis*; that is, the estimation of the necessary values of characteristics such as systemic hierarchy, diameter, vertex connectivity, reliability, cost, and flexibility of topological change, among others. On the other hand, in the topological design next stage of computer networks is performed their *systemic synthesis* which is related to the problem of choosing from a set of given networks a subset that corresponds in the best way to the functions and purposes of the computer network we want to construct while the values of its indices of operation quality keep close to the extreme ones. We will refer to the problem of analysis and synthesis that the network topological design entails.

It is known that in the topological design stage of computer networks, regular networks (regular graphs of degree *k*) with given number of nodes and given diameter, which have a minimal number of edges, are extreme for indices of cost, maximum speed of data transmission, and topological reliability in the case where the technical and geometric characteristics and the cost of nodes (vertices) and transmission channels (edges) are the same [5]. However, since it is not always possible to construct a regular network for all values of a given number of nodes $n$ and given diameter $d$, which has minimal number of edges $m$, some networks with high indices of operation quality should be found among special types of biregular networks.

The search for network structures that meet the practical demands of optimum performance is carried out taking into account the following three fundamental approaches. The first one is related to the search for topologies of networks with a relatively small number of nodes. The second is referred to a generalization and extension of the first and leads to regular topologies of networks whose dimensions depend on the parameters (characteristics) of the quality of network operation. The third allows the design of new networks which, as a general rule, have a large number of nodes and transmission channels and whose construction is carried out using certain operations on existing networks. Such construction procedures are related to the transformation widely used in graph theory called *homeomorphism*. The construction of geodetic networks homeomorphic to a given geodetic network has been discussed in [4]. Such networks satisfy the additional condition that once the transformation of homeomorphism is executed on the given geodetic graph, they continue keeping their property of geodeticity meaning that they keep the structure of geodetic graph which has a great practical design advantage due the fact that geodetic graphs [4] have levels of hierarchy; that is, the same principle of construction on which computer networks are based. Recall that in a geodetic graph, there is a unique shortest path between any two vertices. It is clear that the quality of network operation obtained by using the third approach is mostly determined by the networks obtained by using the first two approaches.

**2. Analysis and Synthesis in Computer Network Topological Design.**

Let us examine the essential topological properties of computer networks which are based on graph theory. In this paper a graph is connected, undirected, and without loops or multiples edges.

Let $G$ be a graph. We define by $\rho(v_i, v_j)$ the length (number of transmission channels) of the shortest path (a geodesic) between nodes $v_i$ and $v_j$. The distance $\rho(v_i) = \max \rho(v_i, v_j)$ is called the diameter of $G$ (of the computer network) at node $v_i$, the magnitude $d = \max \rho(v_i)$ is the diameter of network $G$, and the magnitude $r = \min \rho(v_i)$ is the radius of $G$. The node $v_i$ for which $\rho(v_i) = r$ is called the center of the network [5].

From the point of view of a computer network, the diameter $d$ determines the maximum possible length (the number of transmission channels) of the message transmission route (information packet route) between the source and the recipient provided that the maximum speed of information transmission of the transmission channels does not have any limitations. Since a packet switching is performed in each route node, the magnitude $d$ determines the maximum packet switching number in the route, and as it is expected, the maximum time of packet transmission between the source and the destination.

The design problem of a network for the given topology of nodes consists in choosing the minimum number of transmission channels and their distribution among the nodes so that the maximum network reliability, whose value is not greater than the one that is allowed, can be provided. From the topological point of view, in order to obtain maximum network reliability, the problem is solved under the assumption of equal probability of node operation failure, and also, equal probability of operation failure for the transmission channels.

First of all, given the number of nodes $n$ and transmission channels $m > n$, the degree $deg(v_i)$ is determined taking into account the condition of maximum vertex connectivity for the given network; that is, the condition of the possibility of network construction with a minimum amount of cut nodes of maximum power. The vertex connectivity in the network cannot be greater than min $deg(v_i)$, i.e., we must take into account that

$$\min deg(v_i) \leq [2m/n] = k, \qquad (1)$$

where $[x]$ is the smallest integer greater than or equal to $x$.

Therefore, a network with a high level of reliability (which means a network with possibly at most two paths of shortest length between each pair of its nodes that is called bigeodetic), for which the cost of design is relatively low and the time of information packet exchange between each pair of nodes is minimal, should be found among geodetic and bigeodetic biregular networks of diameter 2 to 5, for which a number $n_0$ of the total of their nodes has degree $k$ and the remaining part $n_1$ has degree $k + 1$. Numbers $n_0$ and $n_1$ are determined from correlation $m = nk / 2$ (that stands for regular networks) which, in the examined case, is reduced to the equation $n_0 k + n_1 (k + 1) = 2m$, provided that $n_0 + n_1 = n$.

In fact, to make the degree of some node less than $k$ means to decrease the vertex connectivity of the network. On the other hand, to increase the degree of some node and make it greater than $k + 1$ means to increase the number of nodes of degree $k$ which increases the minimum number of cut nodes in the network (this is due to the fact that the cut nodes must be among the nodes of minimum degree $k$). Both situations lead to a decrease in the network reliability parameter [5]. Examples of geodetic biregular networks are geodetic graphs homeomorphic to the Petersen graph with diameters $d = 2, 3, 4$ or $5$, which were constructed in [4].

Let us examine some regular constructions of geodetic and bigeodetic networks which can be used in the initial stage of computer network design. According to the ideas previously exposed, one of the most important parameters with which a network can be characterized is its diameter.

**Geodetic graphs of diameter $d = 1$.** This graph is represented by a complete graph.

**Geodetic and bigeodetic graphs of diameter $d = 2$ and $d = 3$.** Fig. 1 exhibits a regular bigeodetic graph for which $n = 8$, $k = 3$ and $d = 2$. The Petersen graph (a geodetic double ring network) is shown in Fig. 2 with parameters $n = 10$, $k = 3$, $d = 2$, $m = 15$. Lastly, Fig. 3 exhibits a bigeodetic double ring network with parameters $n = 16$, $k = 3$, $d = 3$ and $m = 24$.

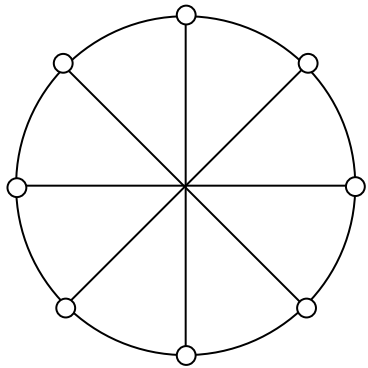

Fig. 1. A regular bigeodetic graph of diameter $d = 2$.

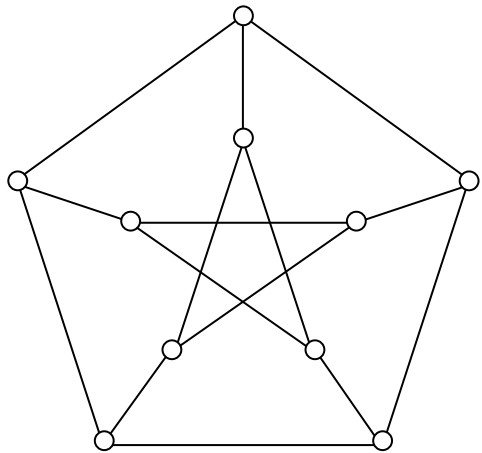

Fig. 2. A geodetic double ring network of diameter $d = 2$.

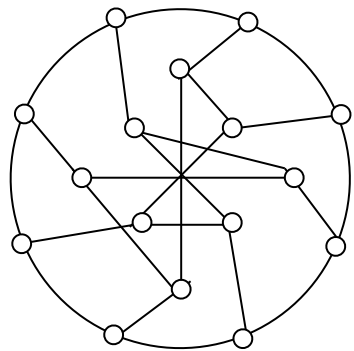

Fig. 3. A bigeodetic double ring network of diameter $d = 3$.

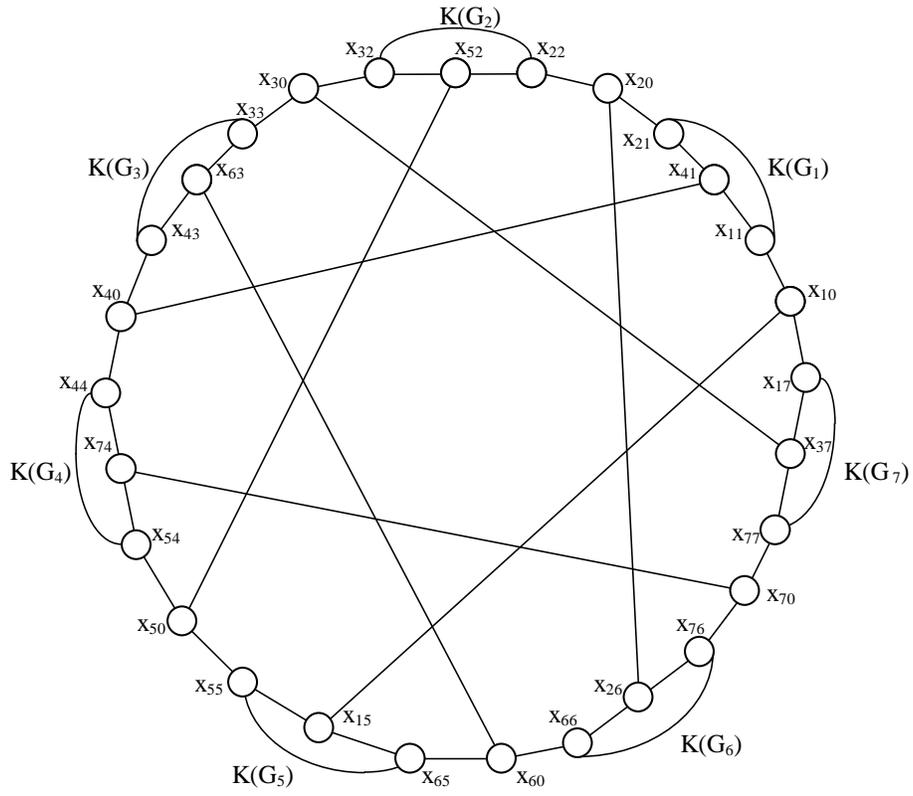

Fig. 4. A geodetic ring network with chords of diameter $d = 4$

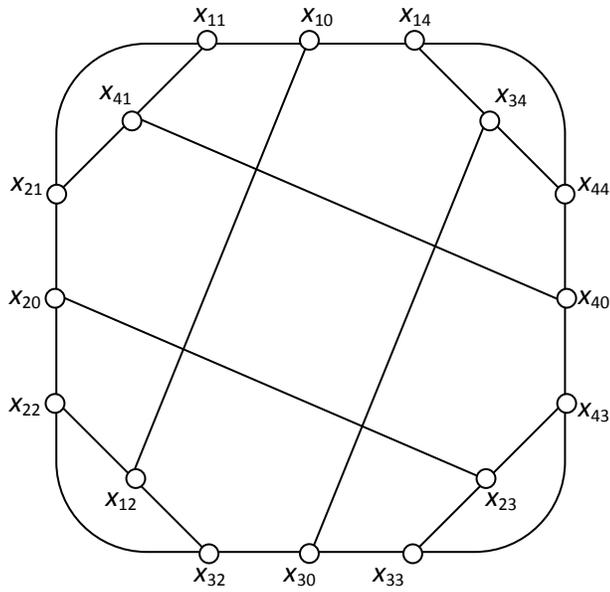

Fig. 5. A bigeodetic ring network with chords of diameter $d = 4$

**Geodetic and bigeodetic graphs of diameter $d = 4$.** Complex 3-regular networks of diameter $d = 4$. Fig. 4 exhibits a 3-regular geodetic ring network with chords of diameter $d = 4$ that has 28 nodes [2] and Fig. 5 exhibits a 3-regular bigeodetic ring network with chords of diameter $d = 4$ that has 16 nodes [3].

The subsequent exposition concerning to characteristics of cost and reliability and how they are related to computer networks and their topology is based on the results obtained in [1]. We need to use the results in [1] to be able to show the real advantages of the practical use of some classes of $k$-geodetic graphs in the topological design of computer networks and perform their topological analysis and synthesis.

The first practical steps in the development of large networks raised questions related to the evaluation of the effectiveness of the problems involved in such a formation, and also, the problem of choosing from a number of alternatives the right topological structure and its cost advantages.

In the topological analysis stage, the problem of associating the cost of the chosen topological structure with the cost characteristics of its component elements arises. It is considered that the best topological structure is either the one that is the most effective for a given budget or the one that has the given parameters at a minimum cost.

In the process of topological design of computer networks, the technical characteristics of nodes and transmission channels are the same. Under this assumption and knowing that the cost of a node is $C_1$ and that of a transmission channel is $C_2$, we will obtain the cost of an [$n, m$]-network with $n$ nodes and $m$ transmission channels:

$$C[n, m] = nC_1 + mC_2. \qquad (2)$$

For a given $n$, a ring ($n$, 2)-network with $n$ nodes, which is 2-regular, is the network of minimum cost while the one of maximum cost is a complete graph ($n, n - 1$):

$$C(n, 2) = n(C_1 + C_2); \; C(n, n - 1) = n[C_1 + ((n - 1)/2)C_2].$$

As a rule, the network design consists in choosing the number of channels and their distribution among the nodes so that it is possible to obtain the required values of reliability and maximum speed of information transmission. It is more convenient to use the so-called relative cost $C[n, m]/(nC_1)$. From the network definition it follows that $m = n + l$ and so the relative cost takes the form $1 + \delta(1 + l/n)$, where $\delta = C_2/C_1$. In this case, as a network cost function it is convenient to use the following expression:

$$A(n, l) = \delta \, l/n. \qquad (3)$$

For a ring we have $l_0 = 0$ and $A_0 = A(n, 0) = 0$ and for a complete graph $l = n(n - 1)/2 - n = n(n - 3)/2$; that is, $A_{max} = A(n, n(n - 3)/2) = \delta(n - 3)/2$. Therefore, for any regular network we have

$$0 \leq A(n, l) \leq \delta (n - 3)/2. \qquad (4)$$

Function $A(n, l)$ can be interpreted as the additional relative cost of a network. The cost of the necessary equipment to operate a network node is made up of the cost of computer equipment and the cost of computer network switching devices. If the cost of the latter is proportional to the number of transmission channels that are connected to a node, then the total cost of computer network switching devices will be proportional to

$$\sum_{i=1}^{n} k_i = 2m$$

and this one can be included in the cost of network transmission channels.

We will examine the asymptotic valuation of cost-effectiveness and its use in choosing the topological structure of a network. The study was performed for configurations such as a bus configuration, complete graphs, double ring networks, and ring networks with chords of length one. The cost of a network is determined by the cost of its nodes and the cost of its transmission channels. The examined topological structures are shown in Fig. 6. We will introduce asymptotic valuations for the general characteristic of cost-effectiveness.

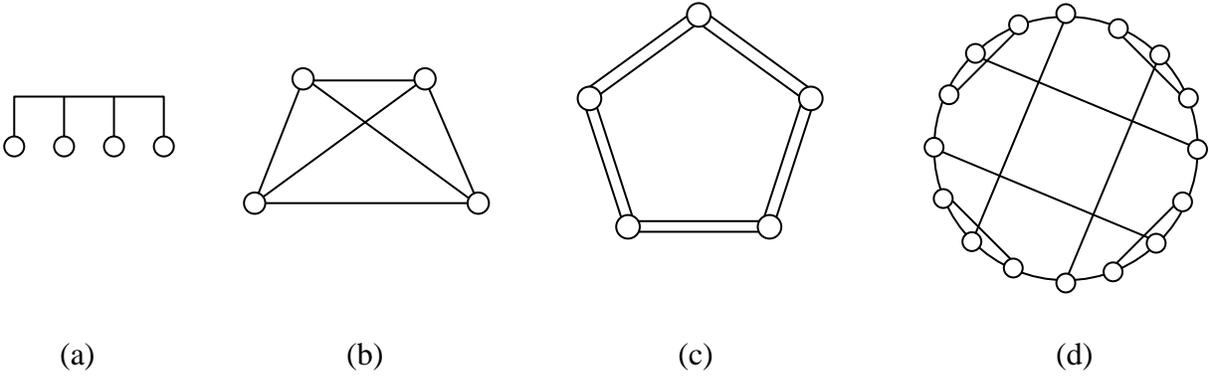

(a)          (b)          (c)          (d)

Fig. 6. Topological structures used for the study of the characteristic of cost-effectiveness.
(a) Bus configuration. (b) Complete graph. (c) Double ring. (d) Ring with chords of length 1.

The cost of each structure is determined as a function of node cost and cost of channels that connect the nodes while the cost of transmission channels does not depend on their length. The calculation of cost gives us the chance to choose, according to the characteristic of cost-effectiveness, the topological structure of the network we want to design. On the other hand, simultaneous transmission and reception of messages is not allowed. The message output speed at one of the nodes is equal to the message input speed at any of the other nodes.

Thus, the cost function of a topological structure can be represented in the form

$$C_R(N_T, n, C_{CP}, C_C, C_{CT}) = nC_{CP} + nkC_C + m_nC_{CT}, \qquad (5)$$

where $C_R$ is the network cost, $N_T$ is the type of topological structure, $n$ is the number of nodes, $C_{CP}$ is the cost of the central and peripheral processors at a node, $C_C$ is the cost of connecting the processors to the computer network switching devices, $C_{CT}$ is the cost of the connection channel of computer network switching devices between two nodes that are connected by this channel, $k$ is the number of transmission channels connected to a node (the degree of the node), and $m_n$ is the number of transmission channels of the topological network that has $n$ nodes.

Each node $i$, in the lapse of a certain time interval $T$, receives and transmits messages over the transmission line. Suppose that the average number of messages, which is observed at a node whose label is $i$, is equal to $Vi$. We will denote the average processing time of messages at node $i$ (both message reception and transmission) by $S_i$, the average message speed at this node by $X_i$, and the probability that the node is busy by $U_i$ (time during which the node is busy with a message). Then we get the following functions.

Attention time at node $i$ is:

$$U_i = X_i\, S_i. \qquad (6)$$

Density of message flow at node $i$ is:

$$k_0 = X_i\,/\,V_i. \qquad (7)$$

So, $$k_0 = U_i\,/\,(V_i\, S_i). \qquad (8)$$

When the number of messages in the network becomes very large, the probability a node is busy approaches 1; that is,

$$\lim_{V_i \to \infty} U_i = 1. \qquad (9)$$

Therefore, $$k_0 \leq 1\,/\,(V_m\, S_m),$$

where $V_m\, S_m = \max V_i\, S_i$. Expression $V_m\, S_m$ can be interpreted as an upper bound of the general time of attention service to messages.

For the sake of simplicity, assume that the processor attention time to a message is $S_{PC}$ and the transmission channel is busy with a message during time interval $S_P$. Considering that the elements of the topological structure are essentially alike, we will assume that each node with label $i$ has the same probability of transmitting a message. Function $k_0$ can be examined as one of the effectiveness measures of the system.

We will determine the required valuations for the examined topological structures. Provided that $n$ is the number of nodes, the average number of calls to each element of a processor at the corresponding node is

$$V_{PC} = 1/n. \qquad (10)$$

Assume an observer wishes to track the working behavior of some node and its $n - 1$ possible receptor ones over the network. Let $R(l, N_T)$ be the number of nodes accessible to an arbitrary node when the data transmission is performed along the topological network of $N_T$ type by the network shortest paths (geodesics) that contain exactly $l$ transmission channels. Then the average number of channels through which the message is sent can be determined by

$$V_{CS} = \frac{\sum_{l=1}^{l_{max}} l \cdot R(l, N_T)}{n-1}, \qquad (11)$$

where $l_{max}$ is the maximum number of channels that can be crossed until the examined node is reached.

The average number of crossings through a concrete transmission channel is

$$V_R = V_{CS}/m. \qquad (12)$$

So, the valuation of $k_0$ is

$$k_0 \leq \frac{1}{\max\{V_{PC}S_{PC}, V_R S_P\}} = \min\left\{\frac{1}{V_{PC}S_{PC}}, \frac{1}{V_R S_P}\right\}. \qquad (13)$$

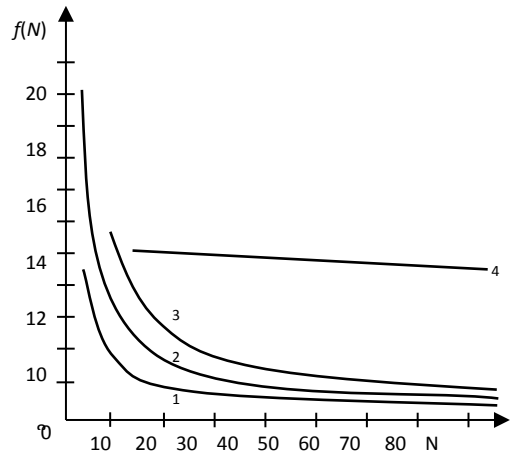

| System's name | Number of nodes | Number of system's connections | Number of channels |
|---|---|---|---|
| Bus Configuration | $n$ | $n$ | 1 |
| Complete Graph | $n$ | $n(n-1)$ | $n(n-1)/2$ |
| Double Ring | $n$ | $4n$ | $2n$ |
| Ring with chords | $n$ | $3n$ | $3n/2$ |

Fig. 7. Functional Dependency $\chi = f(N)$ where $N$ is the number of nodes of the topological structure.
1. Bus configuration. 2. Complete graph. 3. Double ring. 4. Ring with chords ($c = 5$).

We now have the possibility of calculating the asymptotic effectiveness (Figure 7) of the considered structures.

The results of the calculations are shown in Table 1. The elements of general effectiveness of the examined networks are shown.

The choice of the topological network structure is made keeping in mind the principle of cost-effectiveness that has the form

$$\chi = k_0 / C_R = \frac{\min\left\{\dfrac{1}{V_{PC}\, S_{PC}},\, \dfrac{1}{V_R\, S_P}\right\}}{nC_{CP} + nkC_C + m_n C_{CT}}. \qquad (14)$$

Table 1. Elements of effectiveness for some networks ($c$ is the length of the chord in conventional units).

| System | $k_0$ | $V_{PC}\, S_{PC}$ | $V_R^{\max}\, S_P$ |
|---|---|---|---|
| Bus configuration | $1/S_P$ | $S_{PC}/n$ | $S_P$ |
| Complete graph | $n/S_P$ | $S_{PC}/n$ | $2S_P/[n(n-1)]$ |
| Double Ring | $8/S_P$ | $S_{PC}/n$ | $nS_P/[8(n-1)]$, $n$ even; $S_P(n+1)/[8n]$, $n$ odd |
| Ring with chords | $2(c+1)/S_P$ | $S_{PC}/n$ | $\dfrac{2S_P}{n}\cdot\dfrac{n^2}{4(c+1)(n-1)} \leq V_R^{\text{máx}} S_P \leq \dfrac{2S_P}{n}\times \left\{\dfrac{(c+1)\left[\dfrac{n}{2(c+1)}\right]\left(\dfrac{n}{2(c+1)} - \left[2\dfrac{n}{(c+1)}\right] + 1\right)}{n+1} + \left[\dfrac{n}{2(c+1)}\right]\right\}$ |

A concrete calculation requires the values of $S_{PC}$, $S_P$, $C_{CP}$, $C_C$, $C_{CT}$. For topological calculations, we assume that $S_{PC} = S_P$, $C_{CP} = C_C = C_{CT} = 1$.

The given values of conventional costs allow the choice of the topological structure of a complex computer network (Figure 7).

Systems that are technically complex include in their configuration a large number of elements connected in different ways. The complexity of a computer network is negatively reflected in its reliability while the character of the problems to be executed actually demands a high value of the reliability parameter.

$P(t)$ is the fundamental quantitative characteristic of reliability which is determined by the probability of the system non-failure operation; that is, the probability that for the given routines and operation conditions over a period of time $t$, the system will work without failure (without cessation of normal operation or breakdown):

$$P(t) = P\{T > t\},$$

where $T$ is the interval of time during which the system works with no failure, $t$ is a given time, and $P\{A\}$ is the probability that an event $A$ can occur. In this case, the given event $A$ will occur when $T > t$.

By analogy with the reliability function, the non-reliability function (probability of failure) may be introduced:

$$Q(t) = P\{T \leq t\}.$$

From the definition of $P(t)$ and $Q(t)$ it follows that

$$P(t) + Q(t) = 1,$$

which means $$P(t) = 1 - Q(t).$$

For a given interval of time $t$, we get $P = 1 - Q$. From the last correlation it follows that as a measure of $P$, $Q$ can be calculated; that is, the greater the $Q$-value, the worse the valuation of the system reliability.

The network reliability is qualitatively determined by the probability of connection interruption between nodes if the probabilities of transmission line operation failure and node operation failure are known.

In the case of the topological design of computer networks, it is natural to assume that the operation failure of either a transmission channel or a node is a random independent event which is characterized by having equal probability of happening for all nodes ($q_1$) and all transmission lines ($q_2$). Under these conditions, if we calculate the probabilities of failure $Q_{ij}$ between the network nodes $i, j = 1,..., n$, then the network reliability can be determined by the expression

$$\max_{i \neq j} Q_{ij} = Q. \qquad (15)$$

It is of great interest to track the change of probability of failure of an $[n, n + l]$-network as $l$ increases. It is known that for $l = 0$, $Q_2$ is a magnitude of order $(nq_2/2)^2$; that is, $Q_2(0) = \mathbf{O}[(nq_2/2)^2]$, and for $l = 1$, we have $Q_2(1) = \mathbf{O}[2(nq_2/6)^2]$. For $l \ll n$, we can consider that

$$Q_2(l) = \mathbf{O}\left[2\left(\frac{nq_2}{6l}\right)^2\right], \qquad (16)$$

but when $l = n/5$; that is, $n = 5l$ and $m = 6l$, we can consider that

$$Q_2(n/5) = \mathbf{O}[2(q_2)^2], \qquad (17)$$

and that order of probability of failure in the network stays up to $l = n/2$ where it is set to jump to $Q_2(n/2) = \mathbf{O}[2(q_2)^2]$, since for $l = n/2$, $k = [2(n + l)/n] = 3$.

Further, we have

$$Q_2[i(n/2)] = \mathbf{O}[2(q_2)^{i+1}], \; 1 \leq i \leq n - 3. \qquad (18)$$

The change of function $Q_2(l)$ between the jumps at points $n/5$, $n/4$, $n/3$, $n/2$ is not so noticeable. In the initial section of change when $l$ varies from 0 to $n/5$, function $Q_2(l)$ becomes $n^2/8$ times smaller. This explains the sensitiveness to changes in network topology of many heuristic methods that search for networks extreme in reliability and cost which work exactly in this range change (Fig. 8).

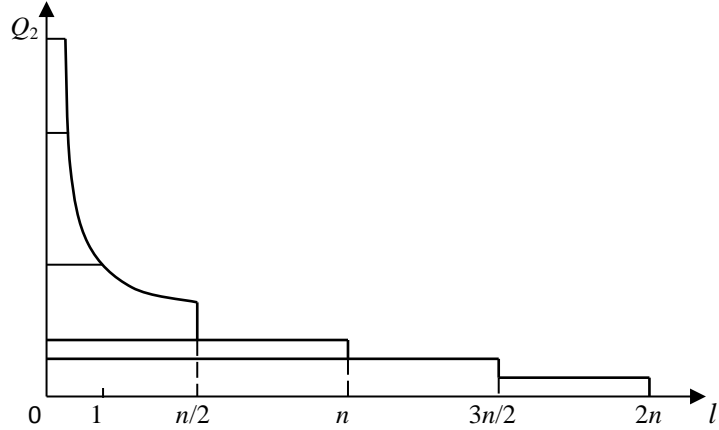

Fig. 8. Reliability change function.

The increase of $l$ leads to the increase of additional expenses in the network organization. If the increase of $l$ is not proportional to the increase of $n$; that is, if the number of nodes $n$ in the network increases while $A(n, l)$ tends to decrease, then $l/n$ must decrease as $n$ increases meaning that $n/l$ should become larger. In this case, the probability of failure in the network also increases according to (16). It is generally accepted that cost savings result in poor performance of optimal systems. However, for the biregular geodetic networks homeomorphic to the Petersen graph with diameters $d = 2, 3, 4, 5$ which were constructed in [4], we have $n = 10, 15, 20, 25$; that is, $l = n/2 = n/3 = n/4 = n/5 =$ constant. So, $l/n$ decreases as $n$ increases, which leads to cost savings, but clearly these networks continue to maintain a high value of network reliability. Table 2 contains the values of reliability and additional relative cost for such geodetic networks. The values of reliability were computed with the help of expression (17) under the assumption that, for topological calculations, the probability of failure of a transmission line is $q_2 = 1/m$.

Table 2. Reliability and additional relative cost for geodetic networks homeomorphic to the Petersen Graph.

| Diameter | Number of nodes | Reliability | Additional Relative Cost $A(n, l)$ |
|---|---|---|---|
| 2 | 10 | 0.9911 | $\delta/2$ |
| 3 | 15 | 0.9950 | $\delta/3$ |
| 4 | 20 | 0.9968 | $\delta/4$ |
| 5 | 25 | 0.9977 | $\delta/5$ |

Note that Table 2 represents the final stage of topological analysis for the geodetic networks homeomorphic to the Petersen graph, which were constructed in [4]. Based on this analysis, it is possible to perform a topological synthesis when choosing the required computer network which is extreme according to parameters of systemic hierarchy (a geodetic network can be organized by levels of hierarchy), diameter (varies from 2 to 5), vertex connectivity (high value since such graphs are blocks whose vertex connectivity is greater than 1), reliability, cost, and also, flexibility of topological change since in [4], different constructions of such geodetic structures having the same number of nodes $n$ and diameter $d$ were generated. The mentioned constructions allow the possibility to change the topology of connections between elements to achieve maximum capability between the required computer network and the functions by it performed.

Lastly, if the additional relative cost $A(n, l)$ gradually increases as $n$ increases, then the networks constructed in [3] using BIB-designs that have a large number of nodes $n$, a high index of reliability (they are bigeodetic blocks), a high level of speed of data transfer (their diameter is 4), and also, a high characteristic of cost-effectiveness (some of them are rings with chords of length one, see Fig. 7) are, from a topological point of view, an alternative of construction of networks with optimal indices of their operation quality.

## 3. Conclusions.

It is especially important to point out that geodetic graphs have application in the mathematical methods of information processing and analysis, in particular, in the general theory of automatic classification also known as cluster analysis, where geodetic graphs are needed in the solution of problems of class structure description. It is known that the information about a class in automatic classification, i.e., the class that is formed by the union of different clusters and its structure, is determined by the choice of parameters of the object description, the given metric, the functional of quality of the automatic classification, and a series of other criteria whose correct formulation and interpretation represent questions that can be solved with the help of the general theory of geodetic graphs. But this problem is object of a more serious study which, for the moment, is not within the limits of this paper and should be examined in a future publication.